# Investigation of photosensitive and photodetector characteristics of n-TPA-IFA/p-Si heterojunction structure

Şükrü Çavdar[1,*], Pınar Oruç[1], Serkan Eymur[2], and Nihat Tuğluoğlu[2]

[1]Faculty of Science, Department of Physics, Gazi University, Ankara, Turkey

[2]Faculty of Engineering, Department of Energy Systems Engineering, Giresun University, Giresun, Turkey

Address correspondence to e-mail: cavdar@gazi.edu.tr

**Abstract**

n-TPA-IFA organic material was synthesized and deposited on p-Si by spin coating method to produce n-TPA-IFA/p-Si heterojunction diode. We determined that the dielectric constant and energy band gap of TPA-IFA organic material were 3.91 and 3.37 eV by DFT/B3LYP/6-311G(d,p) method using on Gaussian 09 W, respectively and the carrier type of TPA-IFA organic semiconductor material was also n-type at room temperature from temperature-dependent hall effect measurements. Using forward and reverse bias $I-V$ measurements in the dark and under various light intensities, we examined the key electrical characteristics of the n-TPA-IFA/p-Si heterojunction diode, including $\Phi_b$ and $n$. It was determined that the rectification ratio (RR) was approximately $10^4$. The reverse current's observed increasing behavior with increasing light indicates that the produced heterojunction diode can be utilized as a photo-diode, detector, or sensor. The photodetector properties of n-TPA-IFA/p-Si heterostructure were explored at different light intensities, and the photoresponsivity ($R$), photosensitivity ($PS$), specific detectivity ($D^*$), and linear dynamic range ($LDR$) of the heterojunction found to be changed with reverse voltage and light intensity. It was found that as light intensity increased, the linear dynamic range (LDR), a crucial characteristic for image sensors, increased as well (10.15 dB and 18.84 dB for 20 and 100 mW/cm$^2$). Ultimately, the findings confirmed that the TPA-IFA-based heterojunction diode could be obtained for the photodetector application.

## 1. Introduction

The potential primary advantages of organic material-based electronic and photoelectrical devices, such as their low cost and easy preparation processes, versatility, and ability to cover large areas, have led to a significant increase in research interest in recent decades. Due to their superior photoconductive capabilities and exceptional thermal and chemical stability, organic semiconducting compounds have been used extensively in photoelectric and electronic devices are used in optoelectronic devices. Organic D-π-A conjugated systems, composed of donor (D) and acceptor (A) groups linked by π-bridges, have received significant attention in various scientific fields due to their unique photophysical properties. Because of their remarkable optical properties, these hybrid D-π-A skeletons have been widely used in materials chemistry for applications such as organic light-emitting diodes (OLEDs), dye-sensitized solar cells (DSSCs), nonlinear optical materials, fluorescence imaging, and memory [1-3].

Thus far, the literature has documented the utilization of organic dyes, specifically D-(π-A), D- π-A, D-D- π-A, D-A- π-A, and (D- π-A). One of the dyes that stands out is the D-A- π-A conjugate system, which offers several significant benefits [4-6]. These include broad absorption and emission bands in the UV-visible region, effective intramolecular charge transfer, easily adjustable molecular orbital energy levels, optoelectronic properties, and narrow bandgap energy. Therefore, utilizing "D-A-π-A" organic sensitizers has been considered a potential strategy for enhancing cell efficiency and optimizing photovoltaic performance in optoelectronic devices, as opposed to conventional "D-π-A" sensitizers.

Organic-inorganic (OI) heterojunctions are one of the electronic and photoelectrical device classes. They are widely studied to use the advantages of both organic and inorganic materials in a single structure [2–20]. A metal/organic material/semiconductor (MOmS) heterojunction is created by adding a thin organic layer in between the metal/Si contacts [7-9]. A thin organic substance between the metal and semiconductor can form a dipole layer, allowing the Schottky barrier height of the junctions to be controlled [10-12]. It has been reported in many articles in the literature that organic semiconductor films on Si semiconductor substrate improve the electronic and optoelectronic properties of metal-semiconductor contacts [13-17].

Triphenylamine (TPA) derivatives, which serve as hole transfer materials, have gained popularity in fabricating electroluminescent devices across multiple industries [18-21]. TPA compounds offer numerous benefits, including their photoconductive and light-emitting characteristics, adjustable electronics, and straightforward processing. Due to their exceptional thermal and optical stability, they are well-suited for altering donor-π-acceptor systems [22-25]. Moreover, TPA derivatives' impressive hole drift mobility enhances their potential for optoelectronic devices [26-30]. Triphenylamine-derived compounds have also been synthesized for several optoelectronic applications. Few of these, however, have had their structural and optical characteristics examined about applying TPA compound as an interfacial layer on metal/p-Si or metal/n-Si thin films. Schiff base compounds have become a topic of interest as organic semiconductors due to their unique electrical and optical properties. Schiff base molecules are often used to create Schottky barrier diodes, but their use is limited due to hydrolysis during device fabrication [31, 32]. The use of Schiff-base compounds in creating Schottky barrier diodes is greatly appreciated due to their straightforward synthesis, low cost, and minimal time requirement. However, while triphenylamine and its derivatives have been used as linkers to create hybrid materials, limited research has been conducted on the impact of TPA-based Schiff base compounds as organic interfacial layers on the electronic parameters of Schottky devices. Therefore, more research on their optical and electrical properties is required to explore the potential applications of TPA-based Schiff-base compounds in optoelectronics.

In this work, we report for the first-time n-type TPA-IFA ("D-A-π-A") compound, in which triphenylamine is used as a donor, the C=N imine part is an acceptor, benzene is a π-spacer, and ester is an acceptor/anchor. Then, n-TPA-IFA was used as a new synthesized organic interfacial layer to modify or improve the photodiode parameters and device performance by generating dipoles at the interface. To the best of the author's knowledge, TPA-IFA thin film has not yet been employed as an interfacial layer in heterojunction and optoelectronic applications in the literature. For this reason, the TPA-IFA organic molecule interface heterojunction devices, n-TPA-IFA/p-Si heterojunction diode, were created, and the optoelectronic and electronic characteristics of the fabricated devices were investigated by using current-voltage (*I-V*) measurements under both different light intensities and dark conditions.

## 2. Experimental
### 2.1 Synthesis and characterization of the TPA-IFA

The synthesis pathway for TPA-IFA is shown in Scheme 1. Compound **2** was synthesized from triphenylamine (**1**) by using Vilsmeier Haack's formylation reaction [33]. To a solution of compound **2** (1.0 mmol) in THF (10 mL), dimethyl 5-aminoisophthalate (3) (2.1 mmol) was added. Then, the mixture was refluxed at 80 °C for 12 h, then the THF was removed via a rotary evaporator. The obtained solid was subjected to filtration, drying, and subsequent recrystallization using ethanol, yielding a pure compound. $^1$H NMR (400 MHz, CDCl$_3$) δ: 8.54 (s, 2H), 8.48 (s, 1H), 8.08-8.06 (m, 4H), 7.64 (d, *J* = 8.5 Hz, 4H), 7.40-7.38 (m, 2H), 7.28-7.21 (m, 5H), 3.99 (s, 12H). $^{13}$C NMR (100 MHz, CDCl$_3$) δ: 166.5, 161.6, 152.0, 149.9, 146.7, 131.4, 130.4, 130.1, 126.7, 126.2, 125.8, 124.2, 52.3.

### 2.2. The device fabrication

Silicon wafer with a 2-inch diameter p-type, orientation (100), 400 μm thickness, 1-10 ohm.cm was cut to suitable sizes with a diamond-tipped pen. First, the p-Si wafer was cleaned with acetone and methanol for 10 and 5 minutes, respectively an ultrasonic bath at 50°C. RCA1 and RCA2 procedures were used for p-Si wafer cleaning. H$_2$O$_2$:H$_2$O solutions were prepared with a mixing ratio of 10:6. The RCA1 cleaning process was used to remove an organic layer; in this process, 5 ml of NH$_3$ (%35) and 25 ml of H$_2$0 was mixed and heated to 70 °C. Then, 5 ml of H$_2$O$_2$:H$_2$O solution was added to this solution. The Si substrate was put in the solution and kept for 15 minutes. RCA2 process was used to remove an inorganic layer; for this cleaning process, 5 ml of HCl (%27) and 25 ml H$_2$O were mixed and heated to 70° C. Then 5 ml of H$_2$O$_2$:H$_2$O solution was added to this solution. The Si substrate was put in the solution and kept for 15 minutes. To remove the oxide layer, the Si wafer was placed in a 2% HF solution for 3 minutes. Then it was rinsed with H$_2$O. The Al ohmic contact was deposited with the help of a tungsten crucible under 10$^{-6}$ bar pressure with a thermal evaporation system for the p-Si wafer. After, the Si wafer was annealed for five minutes at 570 °C in a pure argon atmosphere. Also, a soda-lime glass substrate was cleaned with chloroform, deionized water, methyl alcohol, acetone, and propanol for 10 minutes in an ultrasonic bath for the Hall measurement. 40 mg TPA-IFA was dissolved in 10 ml N,N-Dimethylformamide (DMF), and the solution was mixed for 10 minutes on a magnetic stirrer. A yellowish-clear solution was obtained. TPA-IFA solution was grown on the p-Si wafer and glass substrate by using the spin-coating method at 1000 rpm for 30 seconds. 1.15 mm diameter Al dots contacts were deposited with the same conditions as ohmic contact with a thermal evaporation system for p-Si wafer for electrical *I-V* measurements (Fig. 1). The samples grown on a glass substrate for Hall measurement were cut into a square of 7 mm by 7 mm, and their contacts were done from each corner with the help of silver paste.

Current-voltage measurements were taken between -3V and +3V step by 0.01V at dark and under different light illumination conditions such as 20 mW.cm$^{-2}$, 40 mW.cm$^{-2}$, 60 mW.cm$^{-2}$, 80 mW.cm$^{-2}$, and 100 mW.cm$^{-2}$. Hall measurements were taken by van der Pauw technique at 0.5 T magnetic field between 100K and 400K step by 50K (Fig. 2).

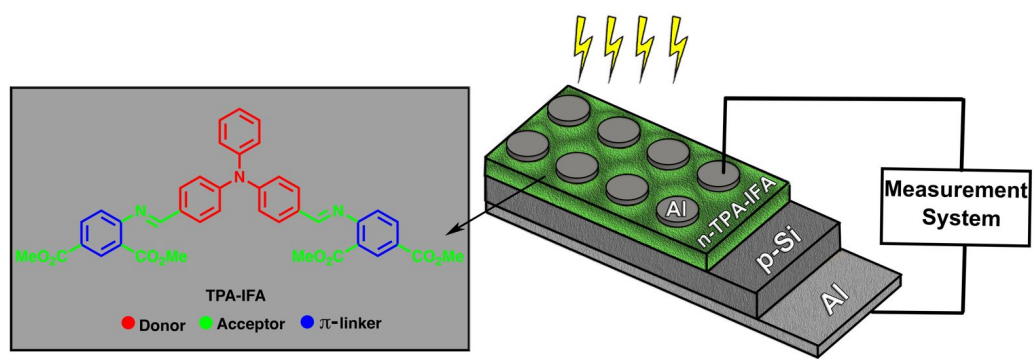

**Fig. 1.** Schematic diagram of n-TPA-IFA/p-Si heterojunction diode

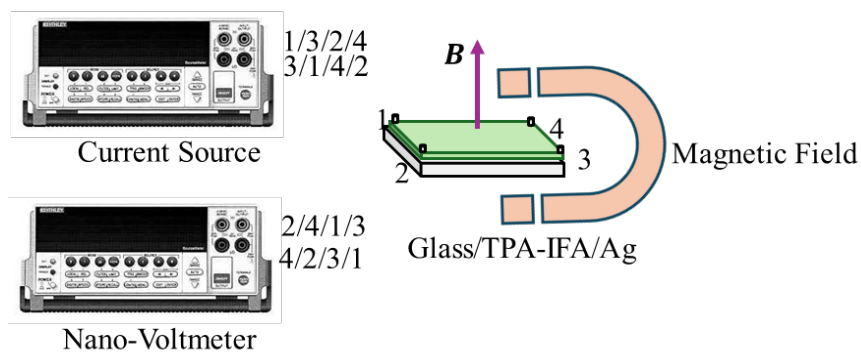

**Fig. 2** Schematic diagram of Hall measurements

## 3. Results and discussion

### 3.1 Theoretical studies

To obtain comprehensive knowledge about the compound's electronic structure, conductivity, and optimal molecular geometry in its ground state, the synthesized TPA-IFA was first computationally evaluated using DFT. The Gaussian 09 program suite software was used to obtain various properties of our compound, including the optimal structural geometry, analysis of the Highest Occupied Molecular Orbital (HOMO) and Lowest Unoccupied Molecular Orbital (LUMO), MEP surface analysis, molecular static isotropic polarizability volume, and molar volume [34]. Figure 3 shows the simulated HOMO and LUMO drawings of the molecule at the gas phase using the DFT-B3LYP/6-311G(d,p) computational level. The computed results of the aforementioned molecular descriptors were given in Table 1 based on these values. As seen, the energy band gap in our molecule was calculated theoretically to be 3.37 eV. As is well known from the literature, ΔE offers details about the compound's conductivity. Since semiconductors' ΔE range is between 0.5 to 3.5 eV, TPA-IFA compound may exhibit semiconducting properties. Moreover, the Clausius-Mossotti equation, which provides a connection between a material's molar volume and polarizability and its dielectric constant, is known to be stated as [35]:

$$\frac{\varepsilon_r - 1}{\varepsilon_r + 2} = \frac{4\pi N_A \alpha'}{3 V_M} \qquad (1)$$

where $\varepsilon_r$ is the dielectric constant, $V_M$ is the molar volume of material, $N_A$ is Avogadro's constant. Also, the $\alpha'$ is the polarizability volume in terms of $\alpha/4\pi\varepsilon_0$. For TPA-IFA at the isolated gas phase state, the compound's dielectric constant was found as 3.91.

**Table 1** Some computed molecular descriptors depending on HOMO and LUMO energy values of the compound.

| Parameters (eV) | Value |
|---|---|
| $E_{\text{LUMO}}$ | -2.207 |
| $E_{\text{HOMO}}$ | -5.577 |
| Energy band gap $|E_{\text{HOMO}} - E_{\text{LUMO}}|$ | 3.370 |
| Ionization potential ($I = -E_{\text{HOMO}}$) | 5.577 |
| Electron affinity ($A = -E_{\text{LUMO}}$) | 2.207 |
| Chemical hardness ($\eta = (I-A)/2$) | 1.685 |
| Chemical softness ($\sigma = 1/2\eta$) | 0.297 |
| Electronegativity ($\chi = (I+A)/2$) | 3.892 |
| Chemical potential ($\mu = -(I+A)/2$) | -3.892 |
| Electrophilicity index ($\omega = \mu^2/2\eta$) | 4.496 |

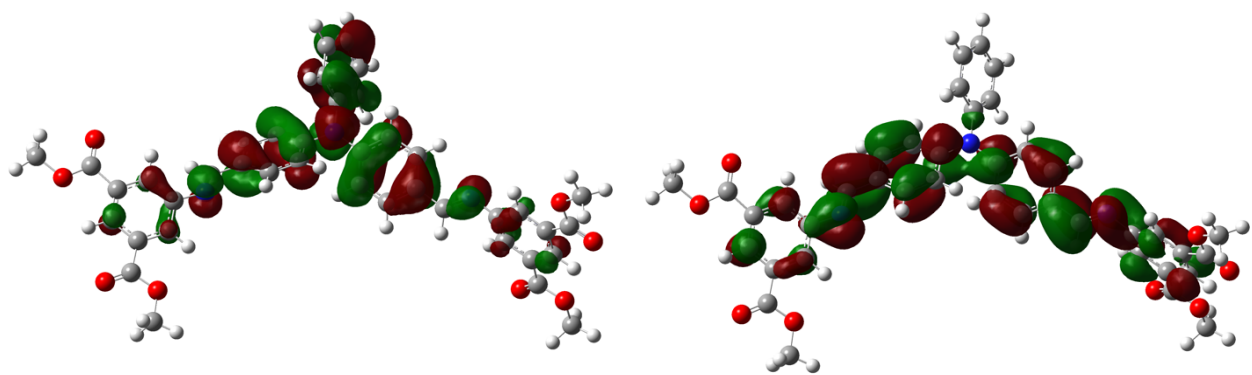

$E_{\text{HOMO}}$ = -5.577 eV  $E_{\text{LUMO}}$ = -2.207 eV

ΔE = |$E_{\text{HOMO}} - E_{\text{LUMO}}$| = 3.370 eV

**Fig. 3** Simulated HOMO and LUMO drawings of the TPA-IFA at the gas phase using the DFT-B3LYP/6-311G(d,p) computational level.

### 3.2 Synthesis of TPA-IFA

Scheme 1 displays the chemical routes used to synthesize the TPA-IFA compound. Firstly, 4,4'-(phenylazanediyl)dibenzaldehyde (**2**) was obtained through Vilsmeier formylation of the triphenylamine [33]. For this, triphenylamine (**1**), which was commercially available, was treated with a POCl3 in DMF for 12 h to give compound **2** in good yield (98%). In a further reaction step, compound **2,** treated with the two equimolar amounts of dimethyl 5-amino isophthalate, formed TPA-IFA as a major product with a yield of 55%.

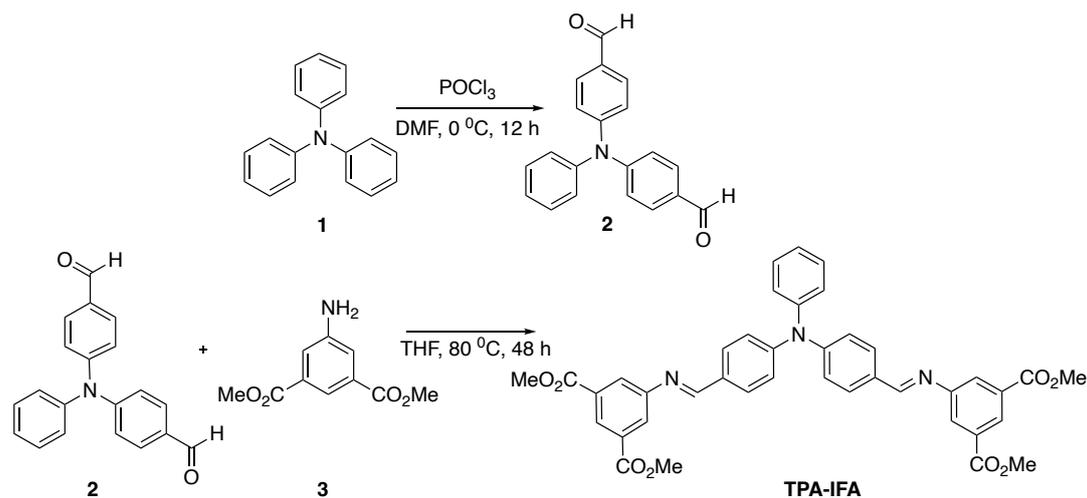

**Scheme 1** Synthetic route for TPA-IFA

The $^1$H NMR spectrum of compound TPA-IFA in CDCl$_3$ is presented in Fig. 4. The structural assignment of TPA-IFA was based on the appearance and disappearance of the $^1$H NMR signals at δ 8.54 for the CH=N protons peaks of the imine groups and at δ 9.88 for the aldehyde protons of compound **2**, respectively. As expected, the characteristic CH$_3$ protons (12H) peaks were observed at 3.99 ppm. In addition, other aromatic ring protons (12H) resonate as singlets at 8.48 (2H) and 8.08 (4H), doublet at 7.64 (d, 4H), and multiplets at 7.38-7.40 (m, 2H), and 7.21-7.28 (m, 5H). In the $^{13}$C NMR spectra of compound TPA-IFA, the signals at 166.5 ppm were assigned to carbons of C=O groups, whereas the imine carbons (C=N-) were observed at 161.6 ppm. The NMR data were in very good agreement with the structure of the TPA-IFA molecule.

### 3.3 The Hall Measurement

Fig. 5 shows a color mapping plot of mobility for TPA-IFA. Hall measurements were taken at 0.5 T magnetic field between 100K and 400K step by 50K. The sample showed p-type characterization at 100K and 150 K, while it showed n-type characterization between 200 K and 400 K. The carrier charge density type of the sample changed between 150K and 200K. The mobility of TPA-IFA varied depending on temperature. The mobility increased with increasing temperature between 100K and 250 K. Then it is decreased for 300K and 350K. Last, it increased to 400K. The temperature with the highest mobility value is 250K. According to these results, the TPA-IFA sample shows n-type behavior at room temperature.

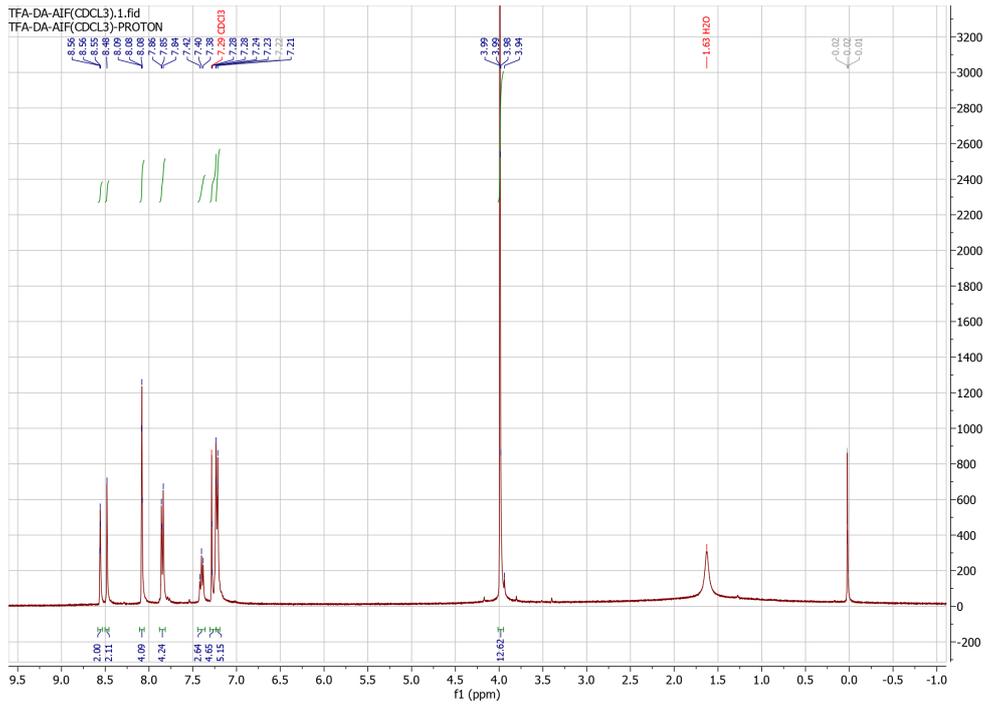

**Fig. 4** $^1$H-NMR spectrum of TPA-IFA

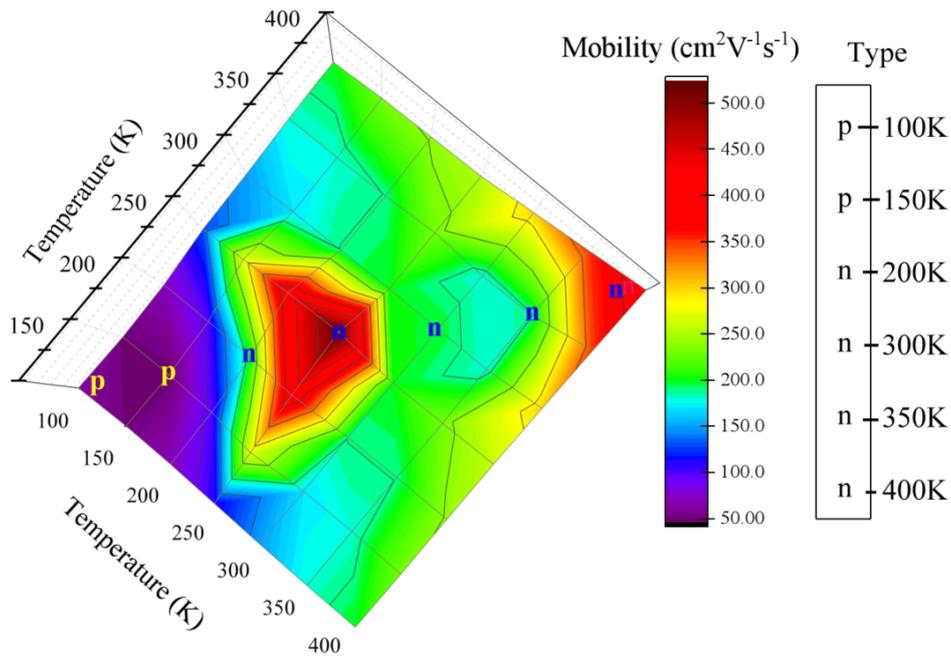

**Fig. 5** Results of hall measurement of the TPA-IFA thin film.

### 3.4. Current-voltage characterization of n-TPA-IFA/p-Si heterojunction

Fig. 7 illustrates the current-voltage (*I-V*) characteristics of the n-TPA-IFA/p-Si heterostructure under dark and different light intensities. The plots clearly demonstrate that all measurement curves exhibit behavior similar to that of a diode. Rectification behavior can be seen from all curves in Fig. 7, and the rectification rate in the dark is very high at $\mp 3$V, and its value is $1.095 \times 10^4$. The rectification rates for dark and different light intensities at $\mp 3$V are given in Table 2. The same graphs additionally demonstrate strong linear behavior at intermediate bias voltages that have distinct slopes, but at sufficiently large forward biases, they deviate from linearity, primarily due to the interface layer. Thermionic emission theory can analyze the *I-V* characteristics of this type of rectifying contact, and the parameters of the heterojunction can be calculated with the help of the following equation (to understand the formation of the n-TPA-IFA/p-Si heterojunction, see Fig. 6): [36-38]:

$$I = I_0 \left[\exp\left(\frac{q(V-IR_s)}{nkT}\right) - 1\right] + \frac{(V-IR_s)}{R_{sh}} \quad (2)$$

where the reverse saturation current ($I_0$) is also given by the following equation [37]:

$$I_0 = AA^*T^2 \exp\left(-\frac{q\Phi_b}{kT}\right) \quad (3)$$

where $A^*$ is the effective Richardson constant, $A$ is the active device area, $n$ is the ideality factor, $\Phi_b$ is the barrier height, $T$ is the temperature, $q$ is the electronic charge, $V$ is the applied voltage, $k$ is the Boltzmann constant, and $R_{sh}$ is the saturation current.

The ideality factor ($n$) and barrier height ($\Phi_b$) values of the fabricated heterojunction diode can be calculated by the following equations [39]:

$$n = \frac{q}{kT}\frac{dV}{d(\ln I)} \quad (4)$$

$$\Phi_b = \frac{kT}{q}\ln\left(\frac{AA^*T^2}{I_0}\right) \quad (5)$$

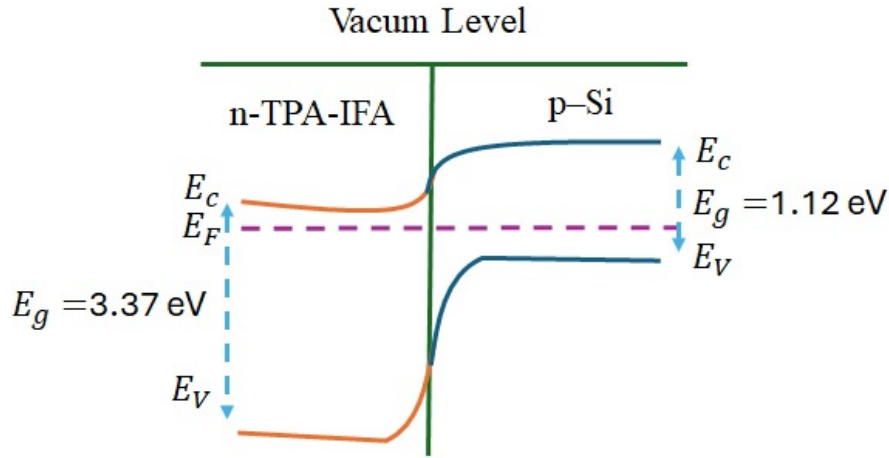

**Fig. 6** Layout of structure formation after contact

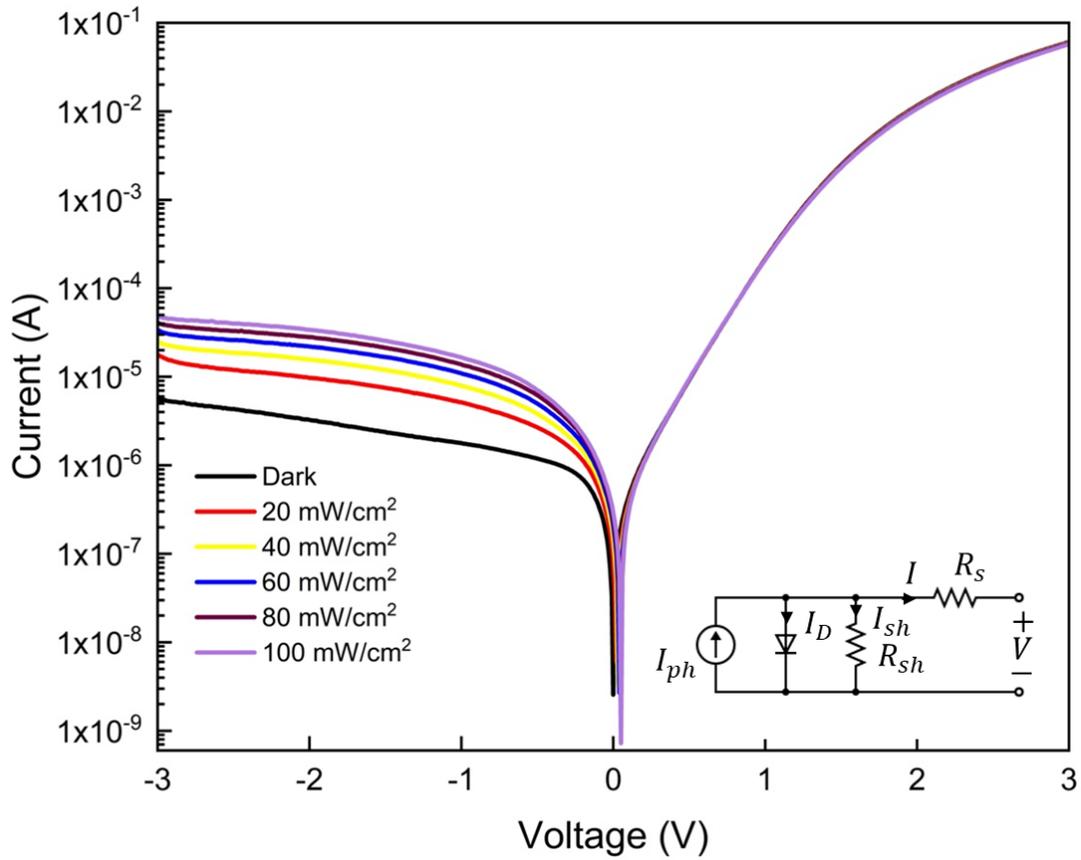

**Fig. 7** The forward and reverse bias ln*I-V* characteristics of n-TPA-IFA/p-Si heterojunction diode in dark and under different light intensities. (inset: schematic diagram of circuit model for n-TPA-IFA/p-Si heterojunction diode)

From thermionic emission (for V≥3kT/q) theory, the intercept and slope of the ln *I-V* plots of Fig. 7 are used to determine the $I_0$, $\Phi_b$ and $n$, respectively. Table 2 lists the computed values of $I_0$, $\Phi_b$ and $n$ values for the n-TPA-IFA/p-Si heterojunction diode in response to varying light intensities and dark. Table 2 illustrates that the barrier height rises with light intensity, and the ideality factor falls comparatively. For the fabricated n-TPA-IFA/p-Si heterojunction diode, the calculated experimental values of $I_0$, $\Phi_b$ and $n$ were 121 nA, 0.716 eV, and 3.01 in the dark and 26.1 nA, 0.755 eV, and 1.63 under 100 mW/cm$^2$. The ideality factor value from Table 2 is greater than the ideal value of 1, indicating that the diode is not behaving ideally. The greater ideality factor value indicates the existence of interface states, an organic TPA-IFA film on the silicon crystal, and series resistance in addition to the inhomogeneities of the Schottky barrier height [40, 41]. As seen in Fig. 7, the photocurrent rises as illumination intensity rises in the reverse bias voltages, confirming that the fabricated heterojunction diode exhibits both photovoltaic and photoconductive activity [42-44]. Also, the circuit diagram of the photodiode is given in Fig. 7 inset. The circuit consists of a parallel shunt resistance ($R_{sh}$)- diode connected in series resistance ($R_s$). Leakage current appeared under light illumination and was shown in the circuit diagram.

**Table 2** The calculated fundamental experimental parameters of the n-TPA-IFA/p-Si heterojunction diode at dark and under different light intensities.

| Light intensity (mW/cm$^2$) | RR x10$^3$ | I-V $I_0$ (nA) | n | $\Phi_b$ (eV) | $R_i - V$ $R_s$ ($\Omega$) | $R_{sh}$ (k$\Omega$) |
|---|---|---|---|---|---|---|
| Dark | 10.95 | 121 | 3.01 | 0.716 | 49.3 | 753 |
| 20 | 3.36 | 82.1 | 2.55 | 0.726 | 49.9 | 211 |
| 40 | 2.29 | 60.6 | 2.24 | 0.734 | 50.6 | 129 |
| 60 | 1.73 | 43.9 | 1.97 | 0.742 | 51.2 | 90 |
| 80 | 1.41 | 36.0 | 1.83 | 0.747 | 51.7 | 69 |
| 100 | 1.18 | 26.1 | 1.63 | 0.755 | 52.2 | 58 |

The resistance found in the metal-interfacial interface is called the series resistance in the heterojunction diode. The heterojunction diode's functionality and its capacity to regulate the flow of electrical current may be impacted by this resistance [29]. Numerous elements, including imperfections in the interfacial layer, metal or semiconductor layer contaminants, and variations in the interfacial layer's thickness and quality, can contribute to series resistance. Multiple ways are available for calculating series resistance in a heterojunction diode [40, 45]. The resistance ($R_i$) values of the heterojunction diode can be determined from Ohm's Law ($R_i = dV_i/dI_i$) both in dark and under different illumination intensities. $R_i$ versus voltage graphs of the n-TPA-IFA/p-Si heterostructure diode for varying the measurement light intensity are shown in Fig. 8. For sufficient lower and higher bias voltages, the values of $R_i$ approach nearly to a constant value; hence, the values of $R_i$ refer to the correct values of the shunt resistance ($R_{sh}$) and series resistance ($R_s$), respectively, both in the dark and under various levels of light. These values were calculated for the fabricated heterojunction diode and are given in Table 2. Because of the increasing current, particularly at reverse biases, the $R_i$ values declined almost regularly as the light intensity increased (Fig. 8, Table 2). For a certain type of light sensor application, this outcome is significant. These values are suitable for a heterojunction diode: 58-753 k$\Omega$ and 49-52 $\Omega$ are the levels of $R_{sh}$ and $R_s$ values.

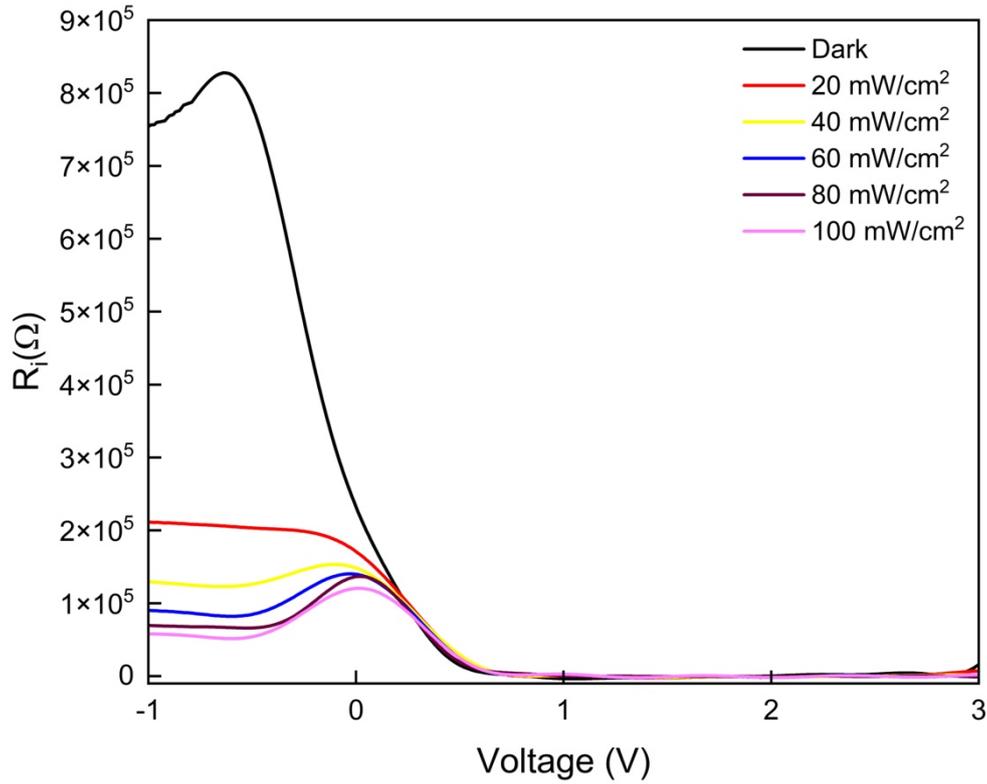

**Fig. 8** $R_i - V$ under different light intensities for the n-TPA-IFA/p-Si heterojunction diode.

Photocurrent ($I_{ph}$), a reverse current, rises with increasing illumination intensities, as shown in Fig. 9. Photodiodes are devices that convert optical signals to electrical current. They have been the subject of extensive research in recent years. Based on light intensity, current measurements verify that the electricity produced photodiode demonstrates photovoltaic activity [46-48]. Through the process of the photovoltaic effect, light energy is transformed into electrical energy by photovoltaic materials. In this context, organic materials such as TPA-IFA thin-film heterojunction can provide several benefits. Photons that penetrate the n-TPA-IFA/p-Si heterojunction diode when it is exposed to light intensity cause electron-hole pairs to form in the semiconductor structures. As a result, a reverse bias voltage is applied to separate the produced electron-hole pairs, and the interface layer's injected holes provide a large photocurrent [49-51]. The following formula yields the photocurrent ($I_{ph}$):

$$I_{ph} = I_{light} - I_{dark} \qquad (6)$$

where $I_{dark}$ and $I_{light}$ represent the reverse bias current values at dark and under different light intensities, respectively. This indicates that because higher energy photons are more likely to overcome the energy barrier, at greater light intensities, the photocurrent likewise increases. The drift velocity of photogenerated electrons and holes at the TPA-IFA/p-Si film interface may cause a rise in photocurrent [52, 53].

Additionally, the interaction between the light intensity ($P$) and the photodiode's photocurrent ($I_{ph}$) is shown below [52]:

$$I_{ph} = AP^m \; ; \quad lnI_{ph} = mlnP + A \qquad (7)$$

where $m$ is an exponent determining the photoconduction process and $A$ is the proportionality constant. The n-TPA-IFA/p-Si heterojunction photodiode's log($I_{ph}$)-log($P$) plot is shown in Fig. 9, and the magnitude $m$ was calculated from its slope, which was 0.621 at –3.0 V. The photoconductive transient mechanism of the n-TPA-IFA/p-Si heterostructure was verified by the linearity seen in Fig. 9, making it a promising option for photodetection applications [54]. The value of the $m$ can be used to identify the nature of the photo-conducting mechanism. If this value falls between 0.5 and 1, it indicates that localized trap levels are continuously distributed [55, 56].

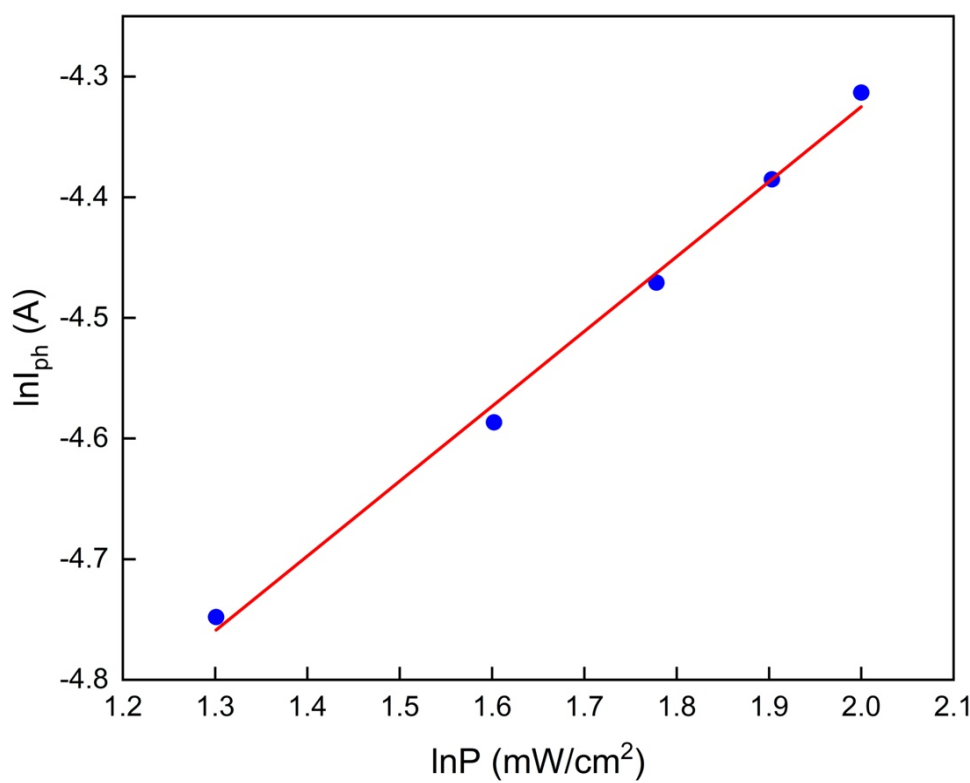

**Fig. 9** log($I_{ph}$)-log($P$) plot for the n-TPA-IFA/p-Si heterojunction photodiode.

Transient photocurrent measurements of the diode were carried out for additional photoresponse analysis, and the results are displayed in Fig. 10. The transient current measurements provide insight into the current conduction mechanism and aid in the determination of several photodetector properties, including photosensitivity, responsivity, and detectivity, for variations in light intensity [50, 57]. As seen in Fig. 10, when the diode is illuminated, electrons contribute to the current, and the quantity of photogenerated charge carriers increases with illumination. Both the diode's current and the quantity of free electrons decrease when the light is turned off. As a result, the trap centers found in the TPA-IFA organic material determine the photoconducting characteristics of the diode.

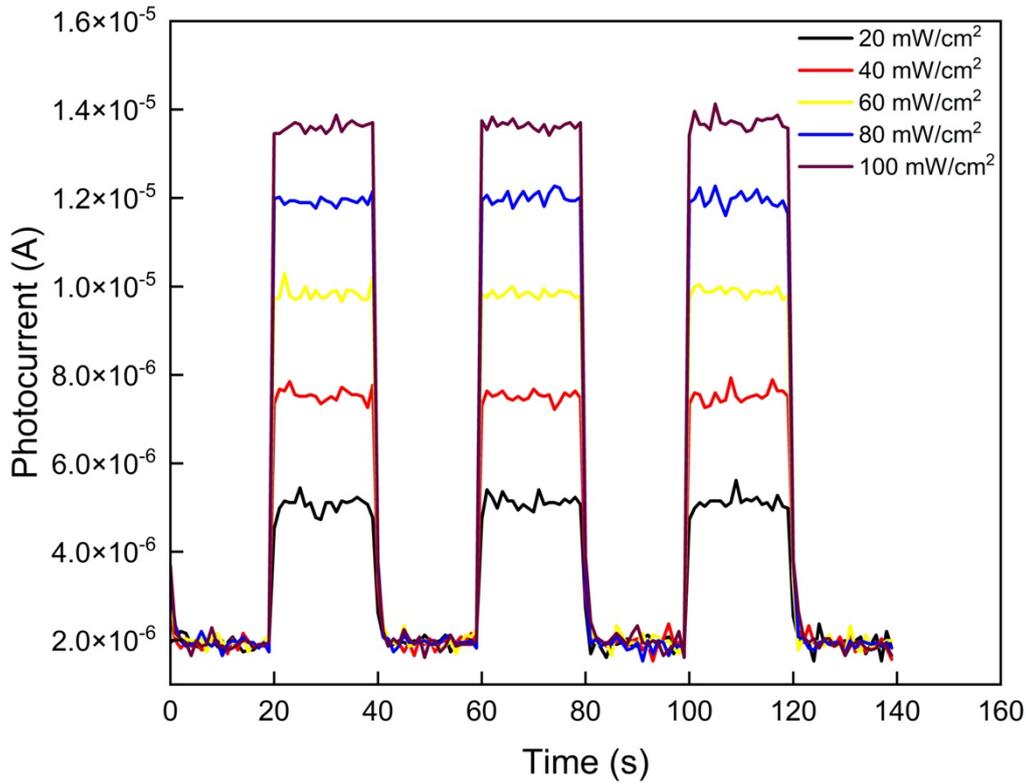

**Fig. 10** Transient photocurrent measurements of n-TPA-IFA/p-Si heterojunction photodiode depending on light intensity

Through the utilization of semiconductors, photovoltaic devices can transform solar energy into electrical energy. Because of the photoelectric effect, these semiconductor structures can take in photons from the sun and then emit electrons as a result of the process. When photons have energies higher than or equal to the energy gap (hv ≥ $E_g$), they can cause additional electrons in the valence band ($E_v$) and trap levels to move to the conduction band ($E_c$) or trap levels, thus absorbing the photons. This process is known as photocurrent ($I_{ph}$). Depending on the intensity of the light, this electron mobility results in the formation of the open-circuit voltage ($V_{oc}$) and short-circuit current ($I_{sc}$) at $I$ =0 A and $V$=0 V, respectively. The characteristics of current density ($J$) against voltage of the n-TPA-IFA/p-Si heterojunction diode are shown in Fig. 11 under different light intensities (from 20 to 100 mW/cm$^2$). The inset of Fig. 11 shows the characteristics of short circuit current density ($J_{sc}$) and open circuit voltage ($V_{oc}$) against the light intensity of the n-TPA-IFA/p-Si heterojunction diode. Also, the calculated values of short circuit current density ($J_{sc}$) and open circuit voltage ($V_{oc}$) with light intensity are illustrated in Table 3. It presents a notable rise in open-circuit voltage ($V_{oc}$) and short-circuit current ($J_{sc}$) values with increasing light intensity, as seen in Fig. 11. This is because a heterojunction diode will absorb more light and produce more photon carriers when the light intensity is increasing. Because the n-TPA-IFA/p-Si structure has photovoltaic characteristics, it can, therefore, be employed as a photodiode. Consequently, the research findings demonstrated that when light intensity increased, the photovoltaic characteristics increased.

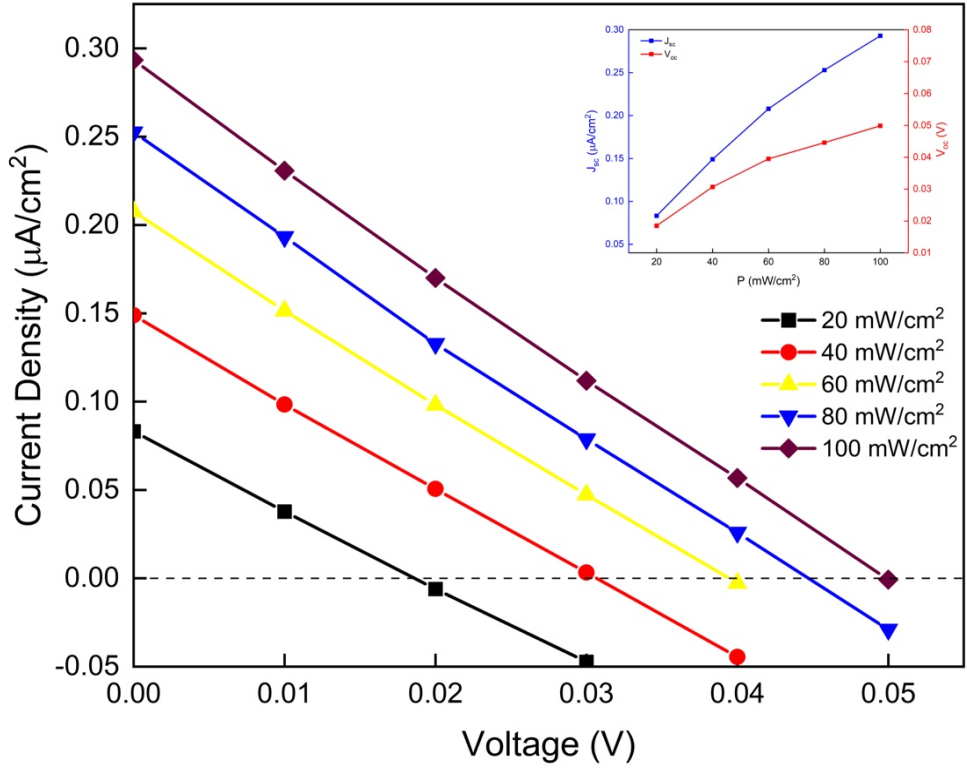

**Fig. 11** Current density-voltage (*J-V*) plots of the n-TPA-IFA/p-Si heterojunction photodiode under different light intensities. The inset shows the changing of $J_{sc}$ and $V_{oc}$ with light intensity (mW/cm$^2$) for the heterojunction photodiode.

The following equations for photoresponsivity (*R*), photosensitivity (*PS*), specific detectivity (*D*\*), and linear dynamic range (*LDR*) were used to determine the detector characteristics of the n-TPA-IFA/p-Si heterostructure diode [58, 59].

$$R = \frac{I_{light} - I_{dark}}{P \cdot A} \tag{8}$$

$$PS(\%) = \frac{I_{light} - I_{dark}}{I_{dark}} \times 100 \tag{9}$$

$$D^* = R\sqrt{\frac{A}{2qI_{dark}}} \tag{10}$$

$$LDR = 20\log\left(\frac{I_{light}}{I_{dark}}\right) \tag{11}$$

Here, *A* represents the detector's active area, and *P* represents the incident light density. One important metric that shows how a photodiode reacts to incoming light is photoconductive photoresponsivity (*R*) [60]. Its definition is as follows: it is the ratio of photocurrent generated to the optical power of incoming light. The variation of *R* for the photodiode with an applied reverse bias voltage at light intensities ranging from 20 to 100 mW.cm$^{-2}$ is depicted in Fig. 12. The change of *R* depending on the light power applied at -3V is given in the inset of Fig. 12. The various values that were found range from 11.20 to 20.59 mA/W, which is advantageous for high-speed diode operation [61, 62]. *R* values depending on light intensity are also given in Table 3.

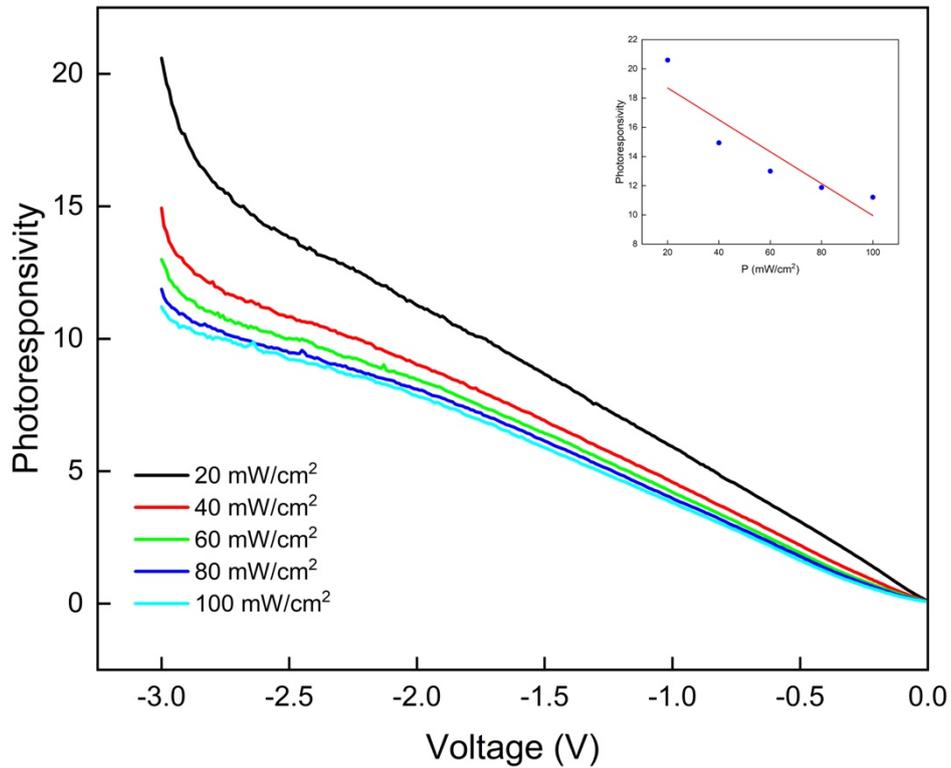

**Fig. 12** Plot of $R$ vs. $V$ of the n-TPA-IFA/p-Si heterojunction diode. The inset shows the plot $R$ vs. $P$ of the heterojunction at -3V.

The ratio of the photocurrent to the dark one for the heterojunction diode is known as photosensitivity ($PS(\%)$). Measurements of photosensitivity characteristics as a function of light intensities are also made to determine whether the device is suitable for use with photodiodes. This crucial variable can also be employed to maintain the level of light-to-current conversion. Figures 13 display the reverse bias voltage profile of the photosensitivity ($PS(\%)$) of the n-TPA-IFA/p-Si heterostructure. The change of $PS(\%)$ depending on the light power ($P$) applied at -3V is given in the inset of Fig. 13. From the inset of Fig. 13, the values of $PS(\%)$ were found to range from 321.76 to 875.44 for 20 and 100 mW/cm², respectively. $PS(\%)$ values depending on light intensity are also given in Table 3.

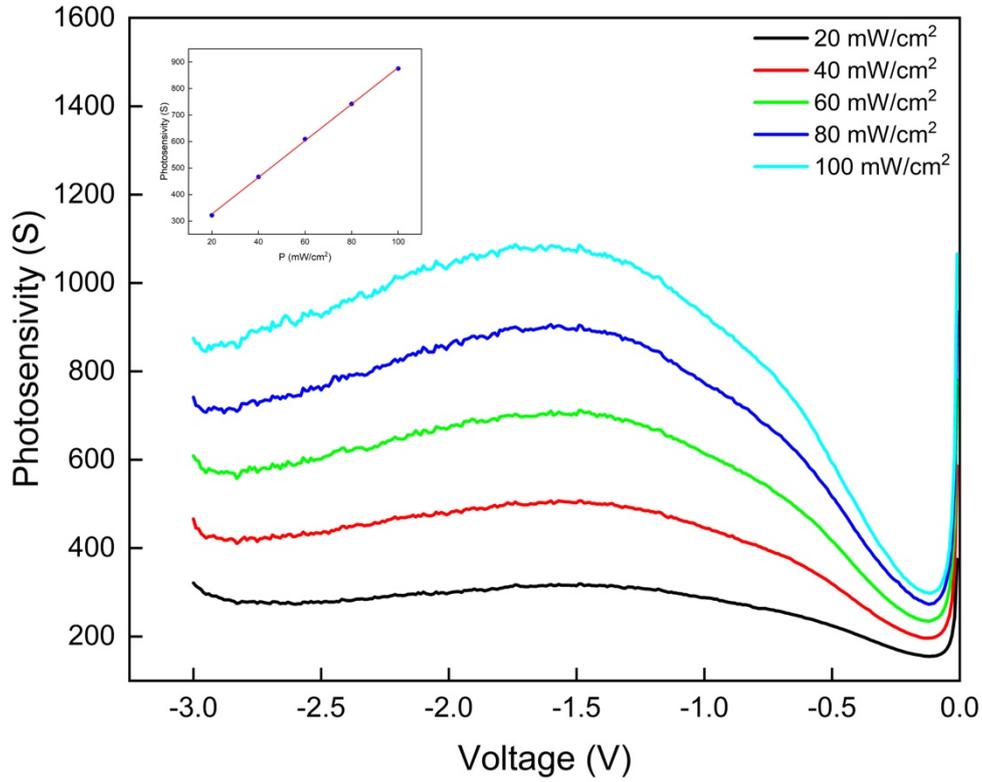

**Fig. 13** Plot of $PS(\%)$ vs. $V$ of the n-TPA-IFA/p-Si heterojunction diode. The inset shows the plot $PS(\%)$ vs. $P$ of the heterojunction at -3V.

The primary parameter for a photodetector is its specific detectivity (D*), which is the minimum light intensity that a diode can detect. In other words, $D^*$ describes a photodetector's capacity to detect low-light signals. This parameter is based on the produced photodiode's noise and responsivity. Figure 14 displays the reverse bias voltage profile of the specific detectivity ($D^*$) of the n-TPA-IFA/p-Si heterostructure. The change of $D^*$ depending on the light power ($P$) applied at -3V is given in the inset of Fig. 14. By increasing the light intensity from 20 to 100 mW/cm$^2$, the computed values of $D^*$ at -3V decreased from $3.22 \times 10^9$ Jones to $1.75 \times 10^9$ Jones. Fig. 18 and the inset of Fig. 18 show how the $D^*$ varies when the reverse bias voltage and light intensity change. Also, $D^*$ values depending on light intensity are also given in Table 3.

The produced n-TPA-IFA/p-Si photodetector's predicted LDR values increase from 10.15 to 18.84 dB when light intensity is raised from 20 to 100 mW/cm$^2$. LDR is essential because image sensors must function over various intensities. For example, if the LDR is sufficiently wide, a clear picture can be created in any situation [63, 64]. LDR values rise as the light intensity rises. Considering the findings, the n-TPA-IFA/p-Si heterojunction diode can be used as an image sensors and photo-devices. LDR values depending on light intensity are also given in Table 3.

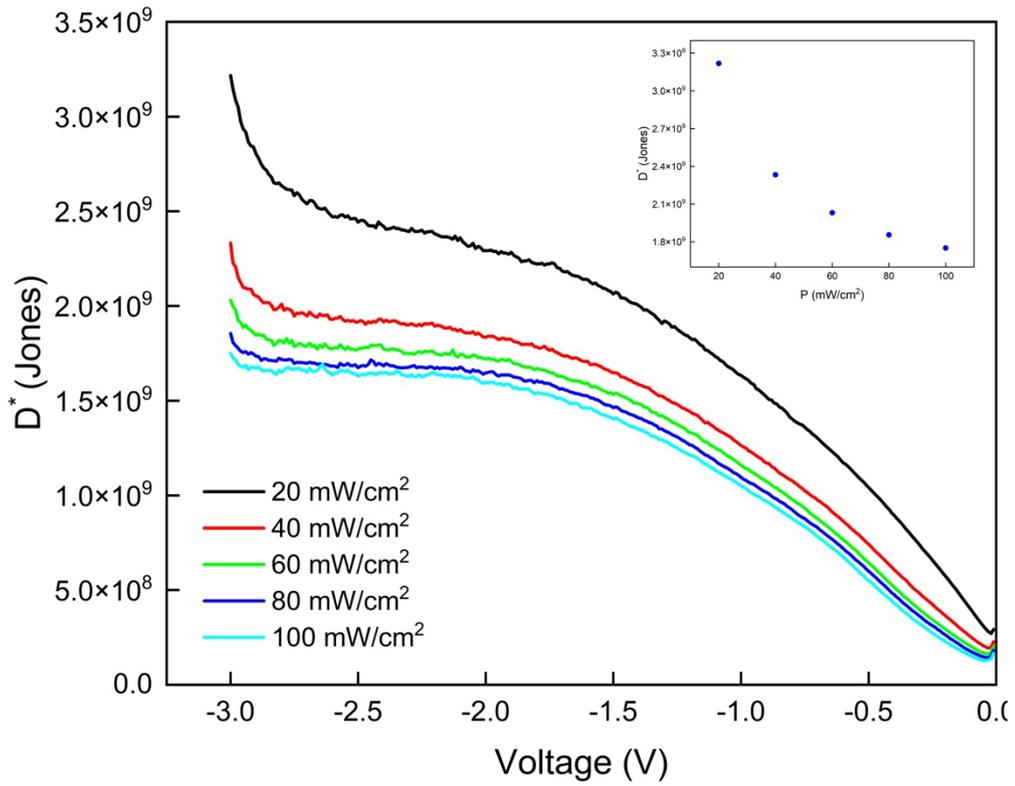

**Fig. 14** Plot of $D^*$ vs. $V$ of the n-TPA-IFA/p-Si heterojunction diode. The inset shows the plot $D^*$ vs. $P$ of the heterojunction at -3V.

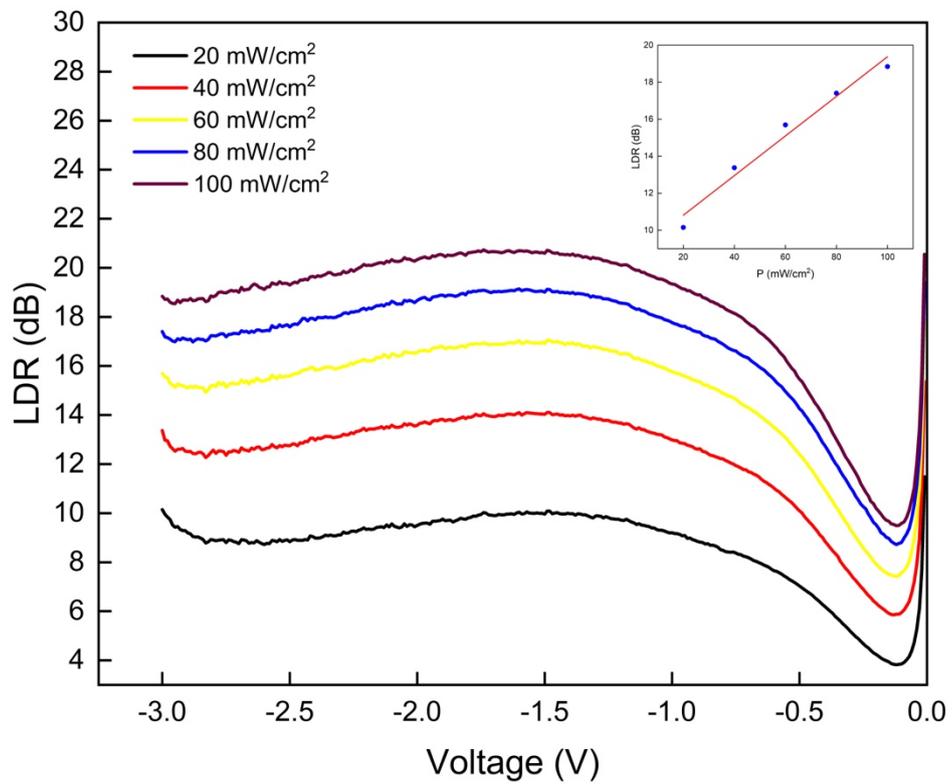

**Fig. 15** Plot of LDR vs. $V$ of the n-TPA-IFA/p-Si heterojunction diode. The inset shows the plot LDR vs. $P$ of the heterojunction at -3V.

**Table 3** Photovoltaic and photodetector parameters of n-TPA-IFA/p-Si heterojunction under different light intensities.

| Power (mW/cm$^2$) | $V_{oc}$ (V) | $J_{sc}$ ($\mu$A) | $R$ | PS(%) | $D^*$ (Jones)x10$^9$ | LDR (dB) | $I_{ph}$ ($\mu$A) |
|---|---|---|---|---|---|---|---|
| | | | | | at $-3.0$ V | | |
| 20 | 0.0185 | 0.083 | 20.60 | 321.77 | 3.22 | 10.15 | 5.14 |
| 40 | 0.0307 | 0.149 | 14.93 | 466.59 | 2.33 | 13.38 | 7.95 |
| 60 | 0.0395 | 0.208 | 12.99 | 609.17 | 2.03 | 15.69 | 10.95 |
| 80 | 0.0446 | 0.253 | 11.87 | 741.97 | 1.86 | 17.41 | 13.79 |
| 100 | 0.0499 | 0.293 | 11.21 | 875.44 | 1.75 | 18.84 | 16.53 |

## 4. Conclusions

In conclusion, we report for the first-time n-type TPA-IFA ("D-A-π-A") compound, in which triphenylamine is used as a donor, C=N imine part is an acceptor, benzene is a π-spacer, and ester is an acceptor/anchor. Also, The HOMO and LUMO energies of the TPA-IFA were calculated by DFT/B3LYP/6-311G(d,p) method using Gaussian 09 W. The hall measurements showed that the TPA-IFA sample shows n-type behavior at room temperature. Then, the spin coating method was used to create a heterojunction device based on an n-type TPA-IFA thin film on the p-type Si substrate. We investigated the photodiode and photodetector parameters of the n-TPA-IFA/p-Si heterojunction diode in both dark and different light conditions, considering its possible optoelectronic applications.

The n-TPA-IFA/p-Si heterojunction diode's ideality factor and Schottky barrier height were found to be, respectively, 0.716 eV and 3.01 in the dark and 0.755 eV and 1.63 in the 100 mW/cm$^2$. In this phenomenon, the ideality factor decreases as the Schottky barrier height increases with increasing light intensity. A heterojunction with a high rectification ratio of 10$^4$ and a low series resistance of 49 Ω was produced. Its lower series resistance value demonstrates the heterostructure's capability for effective photovoltaic cells. A large rectification ratio is seen for heterostructure contact. A decrease in tunneling resistance and space charge modulation across the interface may be the reason for enhanced electrical properties of heterostructures, such as an increase in barrier height and an ideality factor that deviates from unity.

To further investigate the heterojunction diode's response to light, measurements of its photoresponsivity ($R$), photosensitivity ($PS(\%)$), specific detectivity ($D^*$), and linear dynamic range (LDR) were made. The photodiode's $R$, $PS(\%)$, $D^*$, and LDR values were found to be 11.20 mA/W, 321.76, 3.22x10$^9$ Jones, and 10.15 dB at -3 V and 20 mW/cm$^2$, respectively.

According to all the findings, the TPA-IFA organic material is a strong candidate for optoelectronic applications. From the results obtained, it was concluded that the newly synthesized TPA-IFA organic material is a photoactive n-type material and improves device performance, and as a result, it can be used as low-cost optoelectronic devices.

**Acknowledgments**

The authors would like to thank Dr. Sait Malkondu from Giresun University for his help in the synthesis of the TPA-IFA compound. The authors wish to acknowledge the financial support of the Scientific Research Projects Foundation of Gazi University.

**Author Contributions**

ŞC: Investigation, Methodology, Data Analysis, Review and editing, Methodology. PO: Conceptualization, Experimentation, Data Analysis, Writing and editing. SE: Conceptualization, Experimentation, Data Analysis, Writing and editing. NT: Conceptualization, Experimentation, Data Analysis, Writing and editing.

**Funding**

This work was financially supported by the Scientific Research Projects Foundation of Gazi University (BAP-FOA-2022-7389)

**Data availability**

The authors of this study declare that the data that supports their findings is available within the paper. The datasets that were generated and/or analyzed during the study are also available from the corresponding author upon reasonable request.

**Declarations**

**Competing interests** Şükrü Çavdar, Pınar Oruç, Serkan Eymur, and Nihat Tuğluoğlu declare that they have no known competing financial interests or personal relationships that could have appeared to influence the work reported in this paper.